\documentclass[proceedings]{JHEP3}
\usepackage{epsfig}

\PrHEP{jhw2003}
\conference{\raisebox{4pt}{\parbox{11cm}%
{\raggedleft \small
27th Johns Hopkins Workshop on Current Problems in Particle Theory:\\ Symmetries and Mysteries of M Theory}}}

\title{Brane-world Evolution with Brane-bulk Energy Exchange}

\skip\footins = 1\bigskipamount plus 2pt minus 4pt

\newbox\pippobox
\newcommand{\tl}{\tilde t}

\newcommand{\ii}{i}
\newcommand{\jj}{j}
\newcommand{\da}{\dot{a}}
\newcommand{\db}{\dot{b}}
\newcommand{\dn}{\dot{n}}
\newcommand{\dda}{\ddot{a}}
\newcommand{\ddb}{\ddot{b}}
\newcommand{\pa}{a^{\prime}}
\newcommand{\pb}{b^{\prime}}
\newcommand{\pn}{n^{\prime}}
\newcommand{\ppa}{a^{\prime \prime}}
\newcommand{\ppn}{n^{\prime \prime}}
\newcommand{\fda}{\frac{\da}{a}}
\newcommand{\fdb}{\frac{\db}{b}}
\newcommand{\fdn}{\frac{\dn}{n}}
\newcommand{\fdda}{\frac{\dda}{a}}
\newcommand{\fddb}{\frac{\ddb}{b}}
\newcommand{\fpa}{\frac{\pa}{a}}
\newcommand{\fpb}{\frac{\pb}{b}}
\newcommand{\fpn}{\frac{\pn}{n}}
\newcommand{\fppa}{\frac{\ppa}{a}}
\newcommand{\fppn}{\frac{\ppn}{n}}

\title{Brane-world evolution with brane-bulk energy exchange
\footnote{Talk based essentially on \cite{kkttz}}}

\author{Theodore N.~Tomaras  \\
	\llap Department of Physics and Institute of Plasma Physics, University of Crete  
	71003 Heraklion, Greece and \\ Foundation of Research and Technology Hellas \\
	E-mail:	\email{tomaras@physics.uoc.gr}}
 
\abstract{A rich variety of brane cosmologies is obtained once one allows for 
energy exchange between the brane and the bulk, depending on the precise form
of energy transfer, on the equation of state of matter on the brane and on the 
spatial topology. This is demonstrated in the context of a non-factorizable 
background geometry with zero effective cosmological constant on the brane.
An accelerating era is generically a feature
of these solutions. In the case of low-density flat universe more dark
matter than in the conventional FRW picture is predicted, while spatially
compact solutions are found to delay their re-collapse.
In addition to the above, which the interested reader will find 
in greater detail in \cite{kkttz}, a first 
attempt towards a complete description of the full dynamics of both the 
bulk and the brane is reported.}

\begin{document}

\section{Introduction}\label{section1}

The idea that we might be living inside a defect, embedded in a higher
dimensional space has already a long history. Concerning the nature of
the defect, a solitonic codimension one or higher topological object
was proposed~\cite{rubakov1} in the context of an ordinary higher
dimensional gauge field theory, coupled~\cite{shaposhnikov} or not to
gravity. It was soon realized, however, that, in contrast to scalar
and spin-1/2 fields, it would be difficult to confine gauge fields on
such an object. Various interesting ideas and scenaria were
studied~\cite{tetradis}, which have not yet given a fully satisfactory
picture.  In connection with the topology and the size of the bulk
space on the other hand, the popular choice was that the extra
dimensions are compact, with size of order ${\cal O}(M_{Pl}^{-1})$.
It was argued in~\cite{ABLT}, that in the context of the heterotic
string with supersymmetry broken \`a la Scherk-Schwarz this should not
be true anymore.  The scale of supersymmetry breaking is tied to the
size of internal dimensions, and a desirable supersymmetry breaking
scale of a few TeV implies an extra dimension of about
$10^{-16}$cm. Despite difficulties to build a realistic model based on
these ideas, the scenario was taken seriously and analyzed further for
its phenomenological consequences~\cite{benakli}.
 
\looseness=1The situation is drastically different in the context of type-I string
theory~\cite{lykken-witten}. A few developments led to an exciting
possibility and renewed interest in the whole idea.  First, with the
discovery of the D-branes as an essential part of the "spectrum" in
type-I string theory, one could conjecture that we inhabit such a
D-brane embedded in a ten-dimensional bulk. The usual solitonic defect
of field theory was thus replaced by an appropriate collection of
D-branes, which by construction confine the gauge
fields~\cite{polchinski}, together with all the ingredients of the
standard model. All known \pagebreak[3] matter and forces lie on our brane
world~\cite{AKT,AADD}, with the exception of gravity, which acts in
the bulk space as well.  It was, however, pointed out~\cite{ADD}, that
for Kaluza-Klein extra dimensions the gravitational force on the brane
was consistent with all laboratory and astrophysical experimental
data, as long as the extra dimensions were smaller than a
characteristic scale.  This led to the exciting possibility of two
extra dimensions in the sub-millimeter range.  Furthermore, it was
demonstrated in the context of an appropriate effective
five-dimensional theory of gravity, that once we take into account the
back reaction of the brane energy-momentum onto the geometry of
space-time, the graviton is effectively confined on the brane and
Newton's law is reproduced to an excellent accuracy at large
distances, even with a non-compact extra dimension~\cite{rs}.

At the same time, the analysis of the cosmological consequences of the
above hypotheses attracted considerable interest. The first step was
taken in~\cite{binetruy}, where the evolution of perfect fluid matter
on the brane was studied, with no reference to the bulk dynamics and,
therefore, no energy transfer between the brane and the bulk.
Alternatively, a bulk-based point of view was adopted in~\cite{ida},
where the cosmology induced on a moving 3-brane in a static
Schwarzschild-AdS$_5$ background was studied, and a general
interpretation of cosmology on moving branes, together with the idea
of "mirage" cosmology, were presented in~\cite{mirage}.  The
equivalence of the two approaches was demonstrated in~\cite{muko}. It
was also pointed out that branes provide natural mechanisms for a
varying speed of light~\cite{tym}.

Energy-exchange between the brane and the bulk should in principle be
included in any realistic cosmological scenario, and its effects have
been studied in detail in the context of flat compact extra
dimensions~\cite{hall}. The role of energy-exchange on brane cosmology
in the case of non-factorizable extra dimensions has not yet been
investigated extensively~\cite{hebecker,ktt}, even though the
importance of energy outflow from the brane has already been
demonstrated in the context of a Randall- Sundrum configuration with
additional gravity induced on the brane~\cite{ktt}.

Here an attempt towards a more complete analysis of
the cosmic evolution of the brane in the presence of energy flow into
or from the bulk is presented. The aim is to generalize the 
picture in bulk AdS
(bulk gravity plus cosmological constant) by considering a general
bulk theory (that includes gravity).  There may be more bulk fields
and more general bulk-brane couplings. As argued below, one may consider 
a regime where the bulk energy can be
consistently neglected from the equations. Moreover, the energy exchange 
component of the energy-momentum tensor, will be assumed to be
just a power of the matter
density of the brane. Although there are other valid
parameterizations, the one chosen here is motivated by our previous
work in~\cite{ktt}, where we analyzed the out-flow of energy from the
brane due to graviton radiation in the presence of induced
gravity. Standard cross section calculations give an outflow that is
a function of the temperature and other fundamental constants of the
theory.  If one re-expresses the temperature in terms of the running
density using the cosmological equations we obtain a rate of flow that
is a power of the density, with a dimensionful coefficient that
depends on fundamental constants and initial energy densities. This is
valid typically for a whole era, that is, piecewise in the
cosmological evolution.

In a given theory, with a specific bulk content and brane-bulk
couplings, the exponent of the density as well as the coefficient are
calculable functions of the coupling constants of the Lagrangian as
well as initial densities. It is also obvious that if the bulk theory
is approximately conformal the most general form of rate of outflow
will be polynomial in the density. In any case, the analysis carried out here
is "phenomenological", and the parameters used in the energy exchange 
function have not been derived from fundamental dynamics. 

The presentation is organized in five sections of which this introduction is
the first. In section~\ref{section2} the framework of this talk is
described and the approximations on the brane-bulk exchange are
presented. The effective set of equations for the analysis of the brane
cosmology are derived. Section~\ref{section3} contains several
interesting characteristic solutions of the brane cosmology. 
In Section~\ref{section4} a few brief comments will 
be offered concerning the exact treatment of the influx/outflow equations, 
together with a couple of representative cosmological histories, plotted to show 
the rich variety one may expect in the context of this scenario.
Section~\ref{section5} contains a first attempt to describe the full
dynamics of both the bulk and the brane. With appropriate perfect fluids 
in the bulk and on the brane one may describe the late time cosmological
evolution of the brane-world. Our results are summarized and the prospects
for further research along these lines are discussed in the final
section.

\section{The model}\label{section2}

The basic assumption is that we live on a 3-brane embedded
in a 4+1-dimensional bulk with the transverse dimension assumed non-compact.
In particular, we shall be interested in the generic model described by the action
\begin{equation}
S=\int d^5x~ \sqrt{-g} \left( M^3 R -\Lambda +{\cal L}_B^{\rm mat}\right)
+\int d^4 x\sqrt{-\hat g} \,\left( -V+{\cal L}_b^{\rm mat} \right),
\label{001}
\end{equation}
where $R$ is the curvature scalar of the five-dimensional metric
$g_{AB}, A,B=0,1,2,3,5$, $\Lambda$ is the bulk cosmological constant,
and ${\hat g}_{\alpha \beta}$, with $\alpha,\beta=0,1,2,3$, is the
induced metric on the 3-brane. We identify $(x,z)$ with $(x,-z)$,
where $z\equiv x_5$. However, following the conventions of~\cite{rs}
we extend the bulk integration over the entire interval
$(-\infty,\infty)$.  The quantity $V$ includes the brane tension as
well as quantum contributions to the four-dimensional cosmological
constant. ${\cal L}_b^{\rm mat}$ represents the matter on the brane. It 
includes massless excitations (branons) related to fluctuations of 
the brane, which is assumed frozen at $z=0$. Finally, 
${\cal L}_B^{\rm mat}$ stands for the bulk matter action.

As explained in the introduction, various instances of this model
have been extensively discussed in the literature. 
Here we shall present some aspects of the influence of energy 
exchange between the brane and the bulk upon the cosmological 
evolution of our world on the brane. 

We consider an ansatz for the metric of the form
\begin{equation}
ds^{2}=-n^{2}(t,z) dt^{2}+a^{2}(t,z)\gamma_{ij}dx^{i}dx^{j}
+b^{2}(t,z)dz^{2}\,,
\label{metric}
\end{equation}
where $\gamma_{ij}$ is a maximally symmetric 3-dimensional metric.  We
use $k=-1,0,1$ to pa\-ra\-me\-teri\-ze the spatial curvature.

The non-zero components of the five-dimensional Einstein tensor are
\begin{eqnarray}
{G}_{00} &=& 3\left\{ \fda \left( \fda+ \fdb \right) - \frac{n^2}{b^2}
\left(\fppa + \fpa \left( \fpa - \fpb \right) \right) + k
\frac{n^2}{a^2} \right\},
\label{ein00} \\
{G}_{\ii\jj} &=& \frac{a^2}{b^2} \gamma_{ij}\left\{\fpa
\left(\fpa+2\fpn\right)-\fpb\left(\fpn+2\fpa\right)
+2\fppa+\fppn\right\}+
\nonumber \\&&
+\,\frac{a^2}{n^2} \gamma_{ij} \left\{ \fda
\left(-\fda+2\fdn\right)-2\fdda + \fdb \left(-2\fda + \fdn \right) -
\fddb \right\} -k \gamma_{ij},
\label{einij} \\
{G}_{05} &=& 3\left(\fpn \fda + \fpa \fdb - \frac{\dot{a}^{\prime}}{a}
\right),
\label{ein05} \\
{G}_{55} &=& 3\left\{ \fpa \left(\fpa+\fpn \right) - \frac{b^2}{n^2}
\left(\fda \left(\fda-\fdn \right) + \fdda\right) - k
\frac{b^2}{a^2}\right\}.
\label{ein55} 
\end{eqnarray} 
Primes indicate derivatives with respect to $z$, while dots
derivatives with respect to $t$.

\pagebreak[3]

The five-dimensional Einstein equations take the usual form
\begin{equation}
G_{AC} = \frac{1}{2 M^3} T_{AC} \,,
\label{einstein}
\end{equation}
where $T_{AC}$ denotes the total energy-momentum tensor.

Assuming a perfect fluid on the brane and, possibly an additional
energy-momentum $T^A_C|_{m,B}$ in the bulk, we write
\begin{eqnarray}
T^A_{~C}&=& \left. T^A_{~C}\right|_{{\rm v},b}
+\left. T^A_{~C}\right|_{m,b} +\left. T^A_{~C}\right|_{{\rm v},B}
+\left. T^A_{~C}\right|_{m,B}
\label{tmn1} \\
\left. T^A_{~C}\right|_{{\rm v},b}&=&
\frac{\delta(z)}{b}{\rm diag}(-V,-V,-V,-V,0)
\label{tmn2} \\ 
\left. T^A_{~C}\right|_{{\rm v},B}&=&
{\rm diag}(-\Lambda,-\Lambda,-\Lambda,-\Lambda,-\Lambda)
\label{tmn3} \\ 
\left. T^A_{~C}\right|_{{\rm m},b}&=&
\frac{\delta(z)}{b}{\rm diag}(-\rho,p,p,p,0)\,,
\label{tmn4}  
\end{eqnarray}
where $\rho$ and $p$ are the energy density and pressure on the brane,
respectively.  The behavior of $T^A_C|_{m,B}$ is in general
complicated in the presence of flows, but we do not have to specify it
further at this point.

We wish to solve the Einstein equations at the location of the
brane. We indicate by the subscript o the value of various quantities
on the brane.  Integrating equations~(\ref{ein00}), (\ref{einij}) with
respect to $z$ around $z=0$ gives the known jump conditions
\begin{eqnarray}
a_{o^+}'&=&-a_{o^-}'  = -\frac{1}{12M^3} b_o a_o \left( V +\rho \right)
\label{ap0} \\
n'_{o^+}&=&-n_{o^-}' =  \frac{1}{12M^3} b_o n_o \left(- V +2\rho +3 p\right).
\label{np0} 
\end{eqnarray}

The other two Einstein equations~(\ref{ein05}) and~(\ref{ein55}) give
\begin{eqnarray}
\frac{n'_o}{n_o}\frac{\dot a_o}{a_o} +\frac{a'_o}{a_o}\frac{\dot
b_o}{b_o} -\frac{\dot a'_o}{a_o} &=& \frac{1}{6M^3}T_{05}
\label{la1}\\
\frac{a'_o}{a_o}\left( \frac{a'_o}{a_o}+\frac{n'_o}{n_o}\right)
-\frac{b^2_o}{n^2_o}\left( \frac{\dot a_o}{a_o} \left( \frac{\dot
a_o}{a_o}-\frac{\dot n_o}{n_o}\right) +\frac{\ddot a_o}{a_o}\right)
-k\frac{b^2_o}{a^2_o} &=&-\frac{1}{6M^3}\Lambda b^2_o +
\frac{1}{6M^3}T_{55}\,,
\label{la2} 
\end{eqnarray}
where $T_{05}, T_{55}$ are the $05$ and $55$ components of
$T_{AC}|_{m,B}$ evaluated on the brane.
Substituting~(\ref{ap0}), (\ref{np0}) in
equations~(\ref{la1}), (\ref{la2}) one obtains
\begin{eqnarray}
\dot \rho + 3 \frac{\dot a_o}{a_o} (\rho + p) &=& -\frac{2n^2_o}{b_o}
T^0_{~5}
\label{la3}\\
\frac{1}{n^2_o} \Biggl( \frac{\ddot a_o}{a_o} +\left( \frac{\dot
a_o}{a_o} \right)^2 -\frac{\dot a_o}{a_o}\frac{\dot n_o}{n_o}\Biggr)
+\frac{k}{a^2_o} &=&\frac{1}{6M^3} \Bigl(\Lambda + \frac{1}{12M^3} V^2
\Bigr)-
\label{la4} \\&&
-\,\frac{1}{144 M^6} \left( V (3p-\rho ) + \rho (3p +\rho) \right) -
\frac{1}{6M^3}T^5_{~5}\,.
\nonumber
\end{eqnarray}

We are interested in a model that reduces to the Randall-Sundrum
vacuum~\cite{rs} in the absence of matter. In this case, the first
term on the right hand side of equation~(\ref{la4}) vanishes. A new
scale $k_{RS}$ is defined through the relations
$V=-\Lambda/k_{RS}=12M^3 k_{RS}$.

We shall further {\it assume} that the energy exchange between the bulk 
and the brane has negligible effect on the bulk. With this assumption,
which is equivalent to assuming that the brane is embedded in a 
"heat reservoir", 
we shall be able to derive a solution of the brane cosmology that is 
independent fo the bulk dynamics. 

%In order to derive a solution that is largely independent of the bulk
%dynamics, the $T^5_{~5}$ term on the right hand side of the same
%equation must be negligible relative to the second one.  This is
%possible if we assume that the diagonal elements of the various
%contributions to the energy-momentum tensor satisfy the schematic
%inequality\footnote{Strictly speaking, the left hand side
%of~(\ref{vacdom}) concerns only the 55 components of the bulk
%contributions to the energy-momentum tensor. The other components do
%not appear in equations~(\ref{la3}), (\ref{la4}) and do not affect the
%cosmological evolution on the brane.}
%\begin{equation}
%\left| \frac{\left. T\right|^{\rm diag}_{{\rm m},B}}{
%\left. T\right|^{\rm diag}_{{\rm v},B}} \right| \ll \left|
%\frac{\left. T\right|^{\rm diag}_{{\rm m},b}}{ \left. T\right|^{\rm
%diag}_{{\rm v},b}} \right|.
%\label{vacdom} 
%\end{equation}
%Our assumption is that the bulk matter relative to the bulk vacuum
%energy is much less important than the brane matter relative to the
%brane vacuum energy. In this case the bulk is largely unperturbed by
%the exchange of energy with the brane.  When the off-diagonal term
%$T^0_{~5}$ is of the same order of magnitude or smaller than the
%diagonal ones, the inequality~(\ref{vacdom}) implies $T\ll \rho
%k_{RS}$.

At this point we find it convenient to employ a coordinate frame in
which $b_o=n_o=1$ in the above equations. This can be achieved by
using Gauss normal coordinates with $b(t,z)=1$, and by going to the
temporal gauge on the brane with $n_o=1.$ The assumptions for the form
of the energy-momentum tensor are then specific to this
frame.
\footnote{If the vacuum energy dominates over the matter content
of the bulk, we expect that the form of the metric will be close to
the Randall-Sundrum solution with a static bulk.  Thus, we expect
(even though we cannot demonstrate explicitly without a full solution
in the bulk) that in a generic frame, in which
\begin{equation} 
\left| \frac{\left. T\right|^{\rm diag}_{{\rm m},B}}{
\left. T\right|^{\rm diag}_{{\rm v},B}} \right| \ll 1
\label{vacdomm}  
\end{equation} 
we shall have $\dot b\simeq 0$.  Then the transformation that sets $b
= 1$ is not expected to modify significantly the energy-momentum
tensor.}

Using $\beta\equiv M^{-6}/144$ and $\gamma\equiv V M^{-6}/144$, and
omitting the subscript o for convenience in the following, we rewrite
equations~(\ref{la3}) and~(\ref{la4}) in the equivalent form
\begin{eqnarray}
\dot\rho+3(1+w)\,{{\dot a}\over a} \, \rho &=& -T
\label{rho}\\
{{{\dot a}^2}\over {a^2}}&=&\beta\rho^2+2\gamma \rho -
{k\over{a^2}}+\chi
\label{a}\\
\dot\chi+4\,{{\dot a}\over
a}\,\chi&=&2\beta\left(\rho+{\gamma\over\beta}\right)T\,, 
\label{chi}
\end{eqnarray}
where $p=w\rho$, $T=2T^0_{~5}$ is the discontinuity of the zero-five
component of the bulk energy-momentum tensor. As mentioned above, with 
an appropriate choice of parameters we have set to zero the effective
cosmological constant $\lambda=(\Lambda+V^2/12M^3)/12M^3$ on the brane,
which should otherwise be added in th eright-hand side of (\ref{a}).

Equations (\ref{rho}), (\ref{a}) and (\ref{chi}), in place of the usual 
FRW ones, supplemented by a set of initial conditions $(\rho_i,a_i,\chi_i)$
or, equivalently, 
$(\rho_i,a_i,{\dot a}_i)$
\footnote{In the present case ($\rho_i$, $a_i$) 
of conventional FRW cosmology are not enough. 
The reason is that the generalized Friedmann
equation~(\ref{a}) (or~(\ref{twoa})) is not a first integral of the
Einstein equations because of the possible energy exchange between the
brane and the bulk.},
describe the dynamical evolution of matter 
on the brane, within 
the assumptions stated above. They have a straightforward interpretation.
The energy density evolution equation (\ref{rho}) is modified by the
presence of the energy exchange term in the right-hand side.
Eq.~(\ref{a}) is the modification of Friedman equation. It has the extra
term quadratic in the energy density, and may be thought of as 
the \emph{definition} of the
auxiliary density $\chi$. With this definition the other two
equations are equivalent to the original system
(\ref{la3}),~(\ref{la4}).  As we will see later on, in the special
case of no-exchange ($T=0$) $\chi$ represents the mirage radiation
reflecting the non-zero Weyl tensor of the bulk.

The second order equation~(\ref{la4}) for the scale factor becomes
\begin{equation}
{{\ddot a}\over a}=-(2+3w)\beta\rho^2-(1+3w)\gamma\rho-\chi\, .
\label{decel}
\end{equation}
Notice that in the special case of $w=1/3$ one may define a new function $\tilde
\chi \equiv \chi+2\gamma\rho$.  The functions $\tilde \chi, \rho$ and
$a$ satisfy equations~(\ref{rho}) to~(\ref{decel}) with $\tilde\chi$
in place of $\chi$ and $\gamma=0$. This should be expected, since for
$w=1/3$ there is no $\gamma$ left in equation~(\ref{la4}).

It is not surprising that the introduction of the bulk and its interaction 
with the brane enriches the possible cosmologies on the latter. The 
interpretation of observations on the brane becomes unavoidably more
complicated, as a consequence of the at best indirect knowledge of the 
evolution of the bulk and of its interaction with brane matter.
This will become clear in the discussion of possible solutions to which 
we turn next.

\section{Special solutions}\label{section3}

Before we embark on the discussion of the general solutions
of equations~(\ref{rho})--(\ref{chi}), 
it is instructive to start with a
few special cases easy to treat analytically and 
whose physical content is more transparent.
The analysis simplifies considerably in the low-density region, defined by 
$\rho \ll \gamma/\beta=V$, the case in which the matter energy density
on the brane is much smaller than the brane cosmological term. 
In this case one may ignore the term $\beta\rho^2$ in
the above equations compared to $\gamma\rho$.  As a result,
equations~(\ref{rho})--(\ref{chi}) reduce to
\begin{eqnarray}
\dot{\rho}+3(1+w)H\rho&=& -T
\label{onea}\\
H^2=\left(\frac{\dot a}{a} \right)^2&=&
2\gamma\rho + \chi -\frac{k}{a^{2}}
\label{twoa}\\
\dot{\chi} + 4 H \chi &=& 2\gamma T\,.
\label{threea} 
\end{eqnarray}

\subsection{Constant acceleration with $w\neq -1$} \label{section3.1}

An interesting feature of this framework is the possible presence of
accelerating cosmological solutions with constant acceleration, even 
for $w\neq -1$, that would be required in the context of the standard 
FRW cosmology. We can have exponential
expansion with constant Hubble parameter $H\equiv H_*$, even if the brane
content is not pure vacuum energy.

Indeed, let us choose for simplicity $k=0$ and search for a solution 
of equations~(\ref{onea})--(\ref{threea}) with constant $H=H_*$. 
It is straightforward to show that in this case the density $\rho$ and
the auxiliary field $\chi$ are also constant
\footnote{Combining (\ref{onea}) and (\ref{threea}) one obtains 
${\dot\chi}+2\gamma\dot\rho+6\gamma(1+w)H_*\rho+4H_*\chi=0$, which,
combined with (\ref{twoa}) gives $\chi=-2\rho/(3\gamma(1+w))$. This, 
upon substitution into (\ref{twoa}) leads to $\rho=$constant.}. 

Denote with a $*$ subscript the constant values of these quantities.
They satisfy
\begin{eqnarray} 
3H(1+w)\rho_* &=& -T(\rho_*) 
\label{fp1} \\ 
H^2_* &=& 2 \gamma \rho_* + \chi_* 
\label{fp2} \\ 
2 H_* \chi_* &=& \gamma T(\rho_*). 
\label{fp3} 
\end{eqnarray} 
It is clear from equation~(\ref{fp1}) that, for positive matter
density on the brane ($\rho >0$), flow of energy into the brane
($T(\rho)<0$) is necessary.
The accretion of energy from the bulk depends on the dynamical
mechanism that localizes particles on the brane. It is not difficult to
imagine scenaria that would lead to accretion.  If the brane initially
has very low energy density, energy can by transferred onto it by bulk
particles such as gravitons.  An equilibrium is expected to set in if
the brane energy density reaches a limiting value. As a result, a
physically motivated behavior for the function $T(\rho)$ is to be
negative for small $\rho$ and cross zero towards positive values for
larger densities.  In the case of accretion it is also natural to
expect that the energy transfer approaches a negative constant value
for $\rho \to 0$.

The solution of equations~(\ref{fp1})--(\ref{fp3}) satisfies
\begin{eqnarray} 
T(\rho_*) &=& -\frac{3\sqrt{\gamma}}{\sqrt{2}} (1+w)(1-3w)^{1/2}\rho_*^{3/2} 
\label{sol1} \\ 
H_*^2 &=& \frac{1-3w}{2} \gamma \rho_* 
\label{sol2} \\ 
\chi_* &=& - \frac{3(1+w)}{2} \gamma \rho_*\,. 
\label{sol3} 
\end{eqnarray} 
What is the interpretation of this result? 
For a general form of $T(\rho)$, determined by the physics of fundamental
interactions, equation~(\ref{sol1}) is an algebraic
equation with a discrete number of roots. For any value of $w$ in the
region $-1<w<1/3$ a solution is possible. The resulting
cosmological model is characterized by a scale factor that grows 
exponentially with time. The energy density on the brane remains 
constant due to the energy influx from the bulk. The model is very 
similar to the steady state model of cosmology~\cite{steady}. 
The main differences are that the energy density is not spontaneously 
generated, and the Hubble parameter receives an additional contribution 
from the "mirage" field $\chi$ (see equation~(\ref{fp2})).
 
By linearizing equations~(\ref{onea})--(\ref{threea}) around the above 
solution, one may study its stability. The conclusion is that 
for $-1<w<1/3$ and $0<\tilde\nu<3/2$, where
$\tilde\nu\equiv d\ln |T|(\rho_*)/d\ln\rho$, the fixed point solution
described here is stable \cite{kkttz}. 

For $w=-1$ we get the standard inflation only for a value $\rho_*$
that is a zero of $T(\rho)$.  In this case there is no flow along the
fifth dimension and also $\chi_*=0$.

\subsection{"Mirage" radiation for energy outflow} \label{section3.2}
 
Another interesting situation arises in the case of outflow $T>0$, 
of energy from the brane into the bulk and with radiation domination 
on the brane, i.e. with $p=\rho/3$
\footnote{This is the simplest case to treat analytically. The 
conclusions of this subsection generalize \cite{kkttz} to 
other interesting cases, such as dust on the brane.}. 
Equations~(\ref{onea}) and (\ref{threea}) have an exact
solution independent of the explicit form of $T$:
\begin{equation} 
\rho +\frac{\chi}{2\gamma} = \left(\rho_i +
\frac{\chi_i}{2\gamma}\right) \frac{a_i^4}{a^4}
\label{four} 
\end{equation} 
and
\begin{equation} 
H^2= \left(2\gamma \rho_i +\chi_i\right)
\frac{a_i^4}{a^4}-\frac{k}{a^2}\,.  
\label{twoaa} 
\end{equation} 
Assume that initially $\chi_i=0$. It is clear that the effect of the
radiation on the expansion does not disappear even if it decays during
the cosmological evolution: the Hubble parameter of
equation~(\ref{twoaa}) is determined by the initial value of the
energy density, diluted by the expansion in a radiation dominated
universe.  The real radiation energy density $\rho$, however, falls
with time faster than $a^{-4}$.
 
To appreciate the unusual features of the resulting cosmology, consider 
the simple case of what we shall call "radioactive" matter on the brane, 
$T=A\rho$ with $A>0$. Then equation~(\ref{onea}) can be integrated for
arbitrary $w$ to give
\begin{equation} 
\rho=\rho_i \left({a_i \over a} \right)^{3(1+w)} e^{-At}\,,
\label{radioactive} 
\end{equation} 
where we have considered the general case $p=w \rho$.
This equation has an obvious interpretation. The energy density on the
brane dilutes both as a result of the expansion of the universe (the
$a^{-3(1+w)}$ factor) as well as a consequence of its "radioactive decay" 
(the $e^{-At}$ factor). 
In the special case $w=1/3$ we are considering here, one obtains
\footnote{Having
neglected the term $\beta\rho^2$ from equations~(\ref{twoa})
and~(\ref{threea}), these solutions are valid only if
$\beta{\rho}_i^2\ll H_i^2$.  As we shall point out below this 
condition is eventually satisfied for
all solutions in the case of outflow of the form discussed here and
$k=0$.} 
\begin{equation}
\rho=\rho_i \frac{a_i^4}{a^4} e^{-At}\,,
\label{radioactive-a} 
\end{equation} 
\begin{equation} 
\chi = 2 \gamma \rho_i \frac{a_i^4}{a^4} \left( 1-e^{-At} \right).
\label{foura} 
\end{equation}
Imagine an initial energy density $\rho_i\neq 0$, while $\chi_i=0$.
The Hubble parameter, given by equation~(\ref{twoaa}) with $\chi_i=0$,
corresponds to an initial radiation density $\rho_i$, further diluted
only by the expansion. At late times the energy density disappears 
exponentially fast and the expansion is the consequence
of a "mirage" effect, sustained, through the auxiliary $\chi$, by the 
expansion itself. 
Of course, the presence of a "mirage" term is possible even without energy
flow, since equation~(\ref{threea}) has a solution
$\chi=C/a^4$ even for $T=0$, and $\chi$ can act as "mirage"
radiation. The novel feature for $T\not=0$ is that this "mirage"
effect appears through the decay of real brane matter, even if it was
absent in the beginning ($\chi_i=0$).

\subsection{The case of radiation for energy influx} \label{section3.3} 

In the case of radiation the general solution of
equations~(\ref{onea})--(\ref{threea}) was derived in the previous
subsection and is given by equations~(\ref{four}),~(\ref{twoaa}).  The
expansion is that of a radiation-dominated universe with constant
energy $(\chi_i/(2\gamma) + \rho_i)a_i^4$ per co-moving volume.  The
"mirage" energy density is diluted $\sim a^{-4}$.
 
The explicit dependence on time will be discussed next in the case of
flat space ($k=0$), in which the energy density satisfies
\begin{equation} 
\frac{d\rho}{dt} + \frac{2}{t} \rho = -T(\rho)\,.
\label{realen} 
\end{equation} 
If $T(\rho)<0$ for all $\rho$, and the "friction" term in the left
hand side becomes suppressed for $t\to \infty$, we expect an unbounded
increase of $\rho$ in this limit. For $\rho \gtrsim \gamma/\beta$ the
low energy approximation employed in this section breaks down. The
full treatment necessary in this case will be given in the next
section.
  
The actual situation is rather complicated and the details depend
crucially on the form of $T(\rho)$. Assuming that $T(\rho)=A\rho^\nu$
with $A<0$, the exact solution of equation~(\ref{realen}) for
$\nu\not= 1,3/2$ is
\begin{equation} 
\left(\rho \tl^2 \right)^{1-\nu}= \left(\rho_i \tl_i^2 \right)^{1-\nu}
+ \frac{1-\nu}{3-2\nu} \left(\tl^{3-2\nu}-\tl_i^{3-2\nu} \right),
\label{solrad} 
\end{equation} 
where $\tl=|A|t$. For $\nu=1$ the solution is
\begin{equation} 
\rho=\rho_i \frac{\tl_i^2}{\tl^2} e^{\tl-\tl_i},
\label{solrad1} 
\end{equation} 
and for $\nu=3/2$ 
\begin{equation} 
\left(\rho \tl^2 \right)^{-1/2}= \left(\rho_i \tl_i^2 \right)^{-1/2}
-\frac{1}{2}\ln\left(\frac{\tl}{\tl_i} \right).
\label{solrad2} \end{equation} 
 
For $0\leq \nu <1$ we have $\rho\sim \tl^{1/(1-\nu)}$ for $\tl\to
\infty$.  For $\nu=1$ the increase of the energy density at large
$\tl$ is exponential moderated by a power.  For $1<\nu<3/2$ the energy
density diverges at a finite time
\begin{equation} 
\tl^{3-2\nu}_d=\tl^{3-2\nu}_i+\frac{3-2\nu}{1-\nu}\left(\rho_i \tl_i^2
\right)^{1-\nu}\,.  
\label{div} 
\end{equation} 
A similar divergence appears for $\nu=3/2$. For $\nu>3/2$ a divergence
occurs if the quantity
\begin{equation} 
D=\left( \rho_i \tl_i ^2 \right)^{1-\nu} - \frac{\nu-1}{2\nu-3}
\tl_i^{2\nu-3}
\label{quant} 
\end{equation} 
is negative. In the opposite case $\rho \tl^2 \to 1/D$ for $t\to
\infty$, and the energy density diminishes: $\rho \sim t^{-2}$.
 
As we discussed earlier, it is physically reasonable that the energy
influx should stop at a certain value $\rho_{cr}$, and be reversed for
larger energy densities.  The dynamical mechanism that localizes
particles on the brane cannot operate for arbitrarily large energy
densities. This modifies the solutions above that predict an unbounded
increase of the energy density.
 
A final observation that will be encountered again in the next section
is that, despite of the fact that the energy density in most cases
increases for large times, it can decrease at the initial stages.
This is obvious from eq.~(\ref{realen}).  If at the time $t_r$ that
the brane enters a radiation dominated era $|T(\rho)|/\rho<2/t$, the
energy density decreases for a certain time.

\subsection{Non-flat solutions}\label{section3.4}

In addition to the analytical special solutions discussed above, we
would like to present a few suggestive numerical results concerning
the $k=\pm 1$ cases.  For $\nu=1$, we substitute~(\ref{radioactive}),
true for any $A$, into the second order equation~(\ref{la4}), to
obtain
\begin{equation}
\frac{\ddot{a}}{a}+\frac{\dot{a}^{2}}{a^{2}}+\frac{k}{a^{2}}+(1+3w)
\beta\rho_{1}^{2}\,\frac{e^{-2At}}{a^{6(1+w)}}-(1-3w)\gamma\rho_{1}\,
\frac{e^{-At}}{a^{3(1+w)}}=0\,.
\label{back}
\end{equation}
It is obvious from~(\ref{back}) that for outflow with $k=1$ and $w\geq
-1/3$ the Universe will exhibit eternal deceleration.  In particular,
in the case of dust, figure~\ref{fig1} depicts the solution for $a(t)$
of~(\ref{back}) for some initial conditions for $a, \rho$. Notice that
$a(t)$ after a period of decrease, starts increasing again, thus
delaying its re-collapse.  Of course, with appropriate initial
conditions one also obtains solutions with the standard FRW behavior.

\FIGURE[t]{\centerline{\epsfig{file=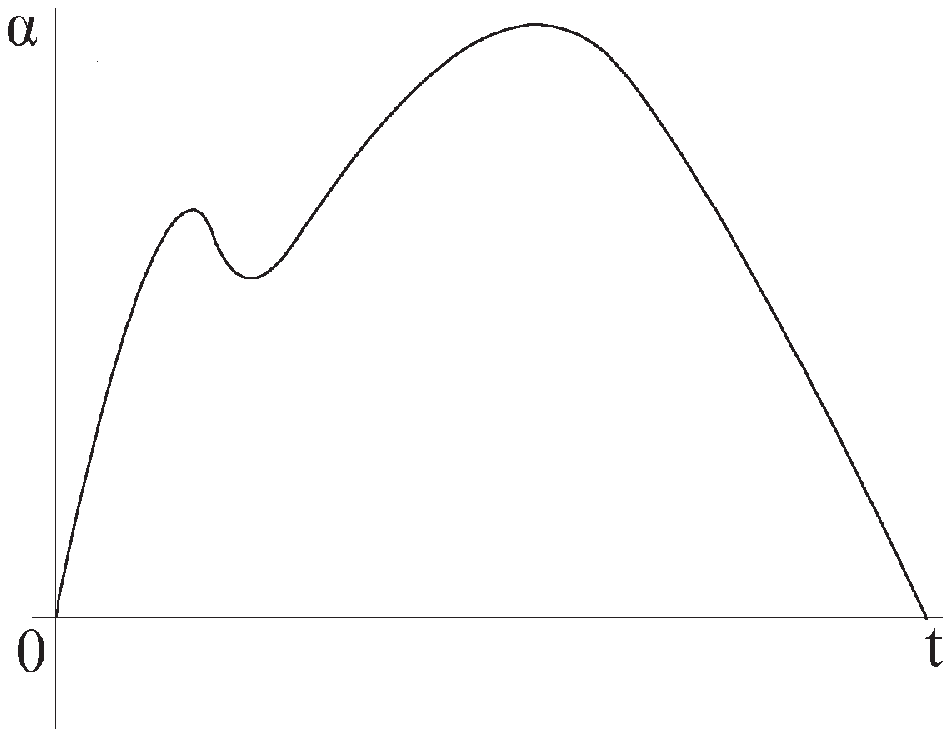,width=.5\textwidth,height=150pt,clip=}}%
\caption{Outflow, $k=+1$, $w=0$, $\nu=1$.\label{fig1}}}

\looseness=-1For an open ($k=-1$) Universe with $A>0$, $w=0$, $\nu=1$ we have found
numerically a solution where $\rho(t)$ monotonically decreases to
zero, while $a(t)$ starts with deceleration, but later on accelerates
eternally; more specifically, $q(t) \rightarrow 0^{+}$ for $t
\rightarrow 0$. Another possibility for $k=-1$ and $A<0$, $w=0$,
$\nu=1$ allows for a Universe starting with acceleration at infinite
densities, which later on, turns to deceleration with $\rho$
approaching a constant value.

\pagebreak[3]

\section{General remarks on the outflow/influx equations}\label{section4}

In \cite{kkttz} the interested reader may find a rather detailed study 
of the fundamental equations of cosmology~(\ref{rho})--(\ref{chi})
in the particular context presented here. 

The general analysis, however, was done numerically for various cases 
of the energy exchange functional dependence on the density $\rho$, various
values of $k$ and various types of matter on the brane.
Two examples with of the resulting cosmologies are the ones shown on 
Figures 2 and 3.

\FIGURE[t]{\centerline{\epsfig{file=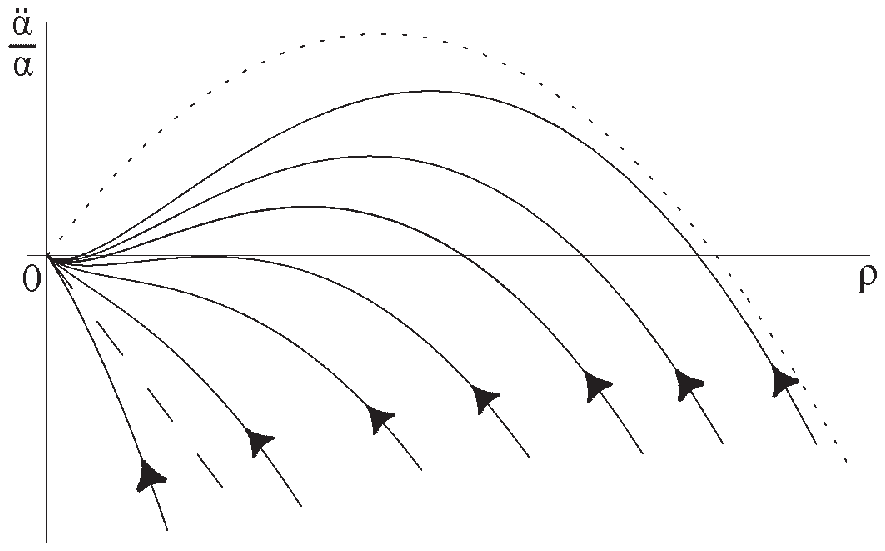,width=.5\textwidth,height=150pt,clip=}}%
\caption{Outflow, $k=0$, $w=0$, $\nu=1$. The arrows show the direction
of increasing scale factor.\label{fig2}}}

Figure 2 shows the global phase portrait of $q\equiv {\ddot a}/a$ versus 
$\rho$ for the case of dust on a flat ($k=0$) brane, with energy {\it outflow} 
and $T=A\rho$. The arrows point in the direction of increasing time, or
equivalently of increasing scale factor for the expanding solutions we
are interested in. In all cases the density decreases with time and the 
universe evolves towards the fixed point with vanishing energy density 
and zero q.  
One may, however, recognize two classes of evolutions. The one in which the
universe decelerates at all times, and the one in which the universe 
first decelerates, then it lives a period of acceleration before it 
decelerates again to approach the fixed point form below the $\rho$ axis.
The behavior near the fixed point may be derived analytically. Also
analytically one may derive the presence of the limiting parabola 
$q=\gamma\rho-\beta\rho^2$, shown with a dotted line. For comparison, 
we also show the straight dashed line $q=-(1+3w)\gamma\rho$, which
represents the standard FRW behavior.

A lot richer and diverse 
is the case of energy {\it influx}. The cosmological evolution of dust 
on a flat brane with linear exchange function $T(\rho)$ is presented 
in Figure 3. Again, as in the previous case, one may recognize 
the limiting parabola, restricting all possible histories of the brane.
Its slope at the origin is also determined analytically and verified
numerically.

\FIGURE[t]{\centerline{\epsfig{file=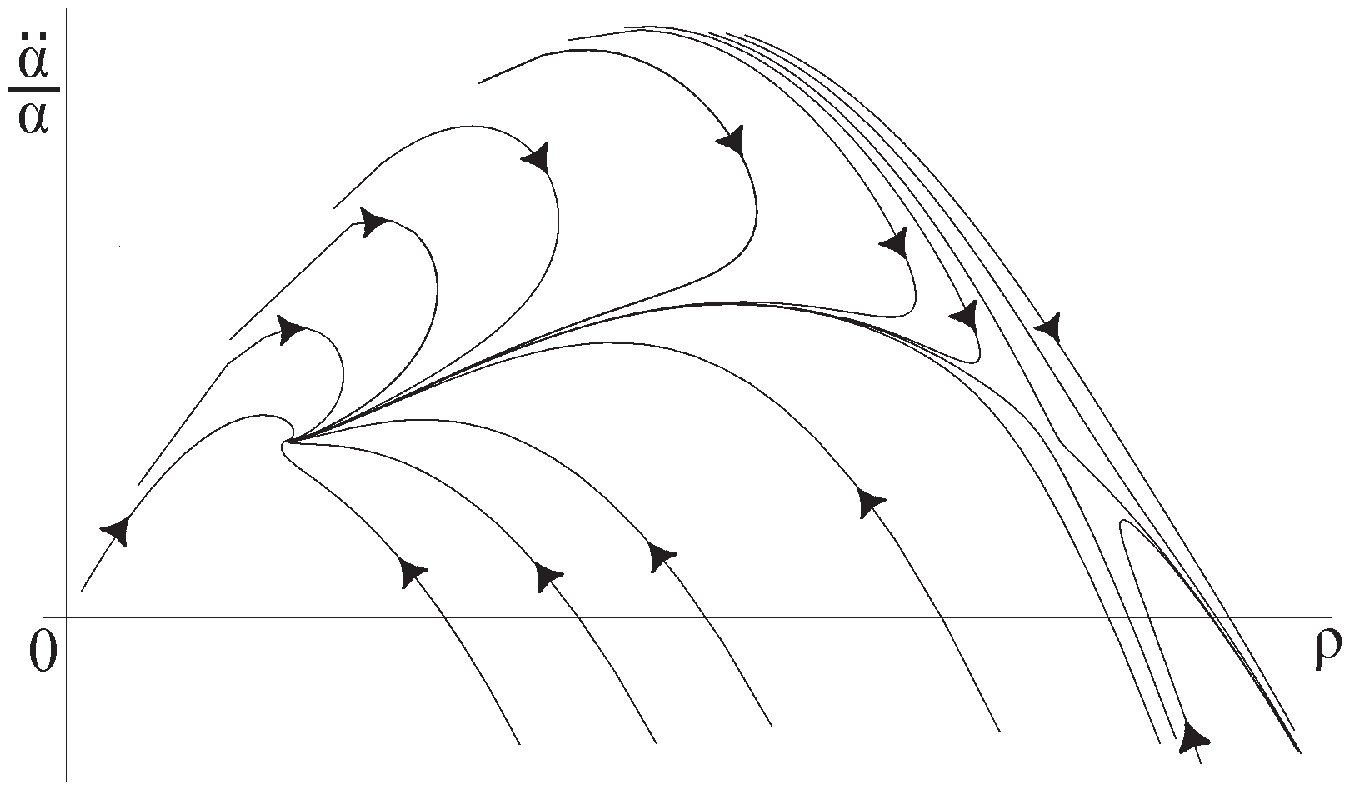,width=.5\textwidth,height=150pt,clip=}}%
\caption{Influx, $k=0$, $w=0$, $\nu=1$.\label{fig3}}}

Not shown is the line $q=-(1+3w)\gamma\rho$ of FRW cosmology.
Due to the energy influx, the origin is not anymore a future fixed point. 
Instead, there are two fixed points, both with constant acceleration
and constant energy density. The one on the left is an attractive fixed
point for almost all initial conditions. The other is a saddle point
corresponding to unstable trajectories. Clearly, there is now a variety 
of possible histories, but it is premature to concentrate on any one of 
them in an effort to understand our Universe. One has to include the 
dynamics of the bulk and solve the complete system, before such an
attempt can be anything but academic.

\section{A first step towards a complete description of the brane-world cosmology.}

The analysis so far was based on the assumption that the bulk is some
kind of reservoir and does not change appreciably during the cosmological
evolution of the brane. So, the bulk was neglected and the autonomous
system of equations for the brane and the energy exchange with the bulk
was discussed. However, realistic applications of this scenario call for
a more general treatment, including the dynamics of the bulk. 

A first step towards a complete analysis of the brane-bulk 
coupled system, is to try at least to describe the stable fixed 
point of the influx case, the far future behavior of the brane-world.  

It is straightforward to convince oneself that the metric
\begin{equation}
ds^2=-dt^2+e^{bt+c|z|}dx^idx^i+dz^2
\end{equation}
satisfies the five-dimensional Einstein equations with
energy-momentum tensor, whose non-vanishing components are 
($8\pi G=1$)
\begin{equation}
T^0_0={3\over 4}(2c^2-b^2)+3\,c\,\delta(z)
\end{equation}
\begin{equation}
T^1_1=T^2_2=T^3_3={3\over 4}(c^2-b^2)+2\, c\, \delta(z)
\end{equation}
\begin{equation}
T^5_5={3\over 4}(c^2-2b^2)
\end{equation}
\begin{equation}
T^0_5={3\over 2}\, b\, c\,\epsilon(z)
\end{equation}

To decide about the nature of matter in the bulk, one has to transform 
to a coordinate frame with vanishing $T^0_5$. This is achieved by a
Lorentz transformation in the $t - z$ subspace. Such a transformation
will leave the $-dt^2+dz^2$ part of the metric invariant, while the 
exponent in the metric will depend linearly on only one of the new
coordinates. 
Specifically, one distinguishes two cases.

Case A. $|c|<|b|$. The transformed metric takes the form
\begin{equation}
ds^2=-dt'^2+e^{b\sqrt{1-c^2/b^2}t'}dx^i dx^i+dz'^2
\end{equation}
with the corresponding energy-momentum tensor
\begin{equation}
{T'}^\mu_\nu=-\frac{3}{4}(b^2-c^2)\;{\rm diag}\,(1,1,1,1,2).
\end{equation}
One interpretation is that in the bulk there exists a perfect 
fluid with constant energy density and pressure, given by
\begin{equation}
\rho_B={3\over 4}(b^2-c^2)>0 \, ,\, P_B=-{3\over 4}(b^2-c^2)=-\rho_B \, ,\, 
P_T={3\over 2}(c^2-b^2)=-2\rho_B
\end{equation}
flowing towards the brane in the brane rest frame.

Case B. $|c|>|b|$. The transformed metric takes the form
\begin{equation}
ds^2=-dt''^2+e^{c\sqrt{1-b^2/c^2}z''}dx^i dx^i+dz''^2
\end{equation}
with the corresponding energy-momentum tensor
\begin{equation}
{T''}^\mu_\nu=-\frac{3}{4}(b^2-c^2)\;{\rm diag}\,(2,1,1,1,1).
\end{equation}
interpreted as a perfect 
fluid with constant energy density and pressure, given by
\begin{equation}
\rho_B={3\over 2}(b^2-c^2)<0 \, ,\, P_B=-{3\over 4}(b^2-c^2)
=-{1\over 2}\rho_B \, ,\, 
P_T=-{3\over 4}(b^2-c^2)=-{1\over 2}\rho_B
\end{equation}
again flowing towards the brane in the brane rest frame. 

The induced metric on the brane is just
\begin{equation}
d{\tilde s}^2=-dt^2+e^{bt}dx^idx^i\; ,
\end{equation}
which for $b>0$ is an exponentially expanding  deSitter space-time 
with constant $q$. 

In addition, on the brane there is a perfect fluid again with constant
energy density and pressure 
\begin{equation}
\rho_b=-3\, c \; , p_b=2\, c=-2\, \rho_b/3. 
\end{equation}
For $c<0$, to avoid exponential blow-up of the metric at transverse-space
infinity, the brane density is positive. Assuming linear state equation for 
the matter on the brane, the situation corresponds to $w_b=-2/3$. 

Finally, there is constant energy influx from the bulk. Indeed, 
$T(z>0)\equiv T^0_5(z>0)/2=3\,b\,c/4<0$, for the chosen signs of $b$ and $c$.

It seems that the above metric provides a complete 
description of the brane-world
cosmology at late times. It realises the main 
characteristics of the stable fixed point 
of Figure 3. It describes a brane carrying 
constant positive density, expanding with constant acceleration 
in the presence of constant influx from the bulk.

There are however a few unsatisfactory features in the above solution. 
First, instead of dust on the brane we had to introduce a fluid with $w=-2/3$. 
Second, assuming a
linear equation of state for the matter in the bulk, 
the above solution of 5-dimensional Einstein gravity with a brane 
does not satisfy in the bulk any of the known energy conditions. 
Indeed, the perfect fluid in the bulk does not
satisfy the null energy condition (NEC) 
($\rho_B+P_B\geq 0$ and $\rho_B+P_T\geq 0$) and consequently, it does not 
satisfy any of the other known energy conditions (weak, strong, dominant or null
dominant). 
Nevertheless, it should be pointed out in connection 
with these comments, that 
on the one hand violation of the energy conditions does not
necessarily mean that the theory is fundamentally sick and unacceptable, 
by being acausal or unstable (Remember, for instance, the recent discussion
in the literature of the so called "phantom" matter). 
On the other hand one does
not have to interpret the above solution assuming a linear relation between 
pressure and density (see for instance \cite{tetra}) and this might allow for
an interpretation with more "natural" matter, both on the brane and in the 
bulk.

\section{Conclusions}\label{section5}

A rather detailed mathematical analysis of the role of the brane-bulk
energy exchange on the evolution of a Brane Universe, adequate for a
wide range of potentially realistic implementations, was presented.
The effective brane \pagebreak[3] cosmological equations were derived with perfect
fluid matter on the brane, constant energy-momentum tensor in the bulk
and non-vanishing exchange between them.  A detailed study of the
solutions of these unconventional equations was performed in the case
of zero effective cosmological constant on the brane, in order to
reduce to the Randall-Sundrum vacuum in the absence of matter.  The
analysis revealed a rather rich variety of possible cosmologies,
depending upon the precise form of the exchange term, the topology of
3-space, and the nature of matter on the brane.
 
\looseness=1 General characteristic 
properties of the solutions were presented, together with 
a few special but particularly interesting solutions, obtained in
the limit of low energy density on the brane.  One of these is the
exactly solvable case of radiation, where a mirage radiation effect
appears through the decay of real brane matter. Another, is a De
Sitter fixed point solution, obtained in the case of energy influx,
even without pure vacuum energy, which in addition is stable for a
wide range of reasonable forms of energy exchange.
 
Clearly, the formulation of a detailed brane-world cosmological scenario 
requires a complete analysis of the full brane-bulk system. A first step towards 
understanding the full dynamics was taken above and a description of 
the asymptotic times cosmology was given. It has a positive brane 
density and a positive q, and represents nicely the 
attactive fixed point plotted in Figure 3, for the case of energy
influx.
Many open questions, such as an all-time description of the brane-world
evolution, the duration of accelerating
periods, the creation of primordial fluctuations, and the
compatibility with conventional cosmology at low energy densities,
should be addressed. However, we believe that the cosmological
evolution in the context presented here has many novel features,
that may provide answers to outstanding questions of modern cosmology.

\acknowledgments
I would like to thank R. Guedens,
A. Kehagias, G. Panotopoulos and N. Tetradis for several 
helpful discussions.  
This work was partially supported by European Union under the RTN
contract HPRN--CT--2000--00122.


\begin{thebibliography}{99} 

\bibitem{kkttz} E. Kiritsis, G. Kofinas, N. Tetradis, T.N. Tomaras and
V. Zarikas, \emph{Cosmological evolution with brane-bulk energy exchange},
JHEP 0302:035,2003 [\hepth{0207060}].

\bibitem{rubakov1}
V.A. Rubakov and M.E. Shaposhnikov, \emph{Do we live inside a domain
  wall?}, \plb{125}{1983}{136}.

\bibitem{shaposhnikov}
See for instance E.~Roessl and M.~Shaposhnikov, \emph{Localizing
gravity on a't~Hooft-polyakov monopole in seven dimensions},
\prd{66}{2002}{084008} [\hepth{0205320}], E. Roessl and
M. Shaposhnikov, and references therein.

\bibitem{tetradis}
A.T. Barnaveli and O.V. Kancheli, \emph{The world as a membrane in a
  space of higher dimensionality}, \sjnp{51}{1990}{573};\\
\emph{The gauge fields on the soliton membrane. (In russian)},
\sjnp{52}{1990}{576};\\
G.R. Dvali and M.A. Shifman, \emph{Domain walls in strongly coupled
   theories}, \plb{396}{1997}{64} [\hepth{9612128}];\\
N.~Tetradis, \emph{The world as a dual josephson junction},
\plb{479}{2000}{265} [\hepph{9908209}];\\
A.~Kehagias and K.~Tamvakis, \emph{Localized gravitons, gauge bosons
  and chiral fermions in smooth spaces generated by a bounce},
  \plb{504}{2001}{38} [\hepth{0010112}].

\bibitem{ABLT}
I.~Antoniadis, C.~Bachas, D.C. Lewellen and T.N. Tomaras, \emph{On
  supersymmetry breaking in superstrings}, \plb{207}{1988}{441};\\
T. Banks and L.J. Dixon, \npb{307}{1988}{93}.

\bibitem{benakli}
I.~Antoniadis, \emph{A possible new dimension at a few TeV},
\plb{246}{1990}{377};\\
I.~Antoniadis and K.~Benakli, \emph{Limits on extra dimensions in
  orbifold compactifications of superstrings}, \plb{326}{1994}{69}
  [\hepth{9310151}];\\
I.~Antoniadis, K.~Benakli and M.~Quiros, \emph{Production of
  Kaluza-Klein states at future colliders}, \plb{331}{1994}{313}
  [\hepph{9403290}];\\
I.~Antoniadis, K.~Benakli and M.~Quiros, \emph{Direct collider
  signatures of large extra dimensions}, \plb{460}{1999}{176}
  [\hepph{9905311}].

\bibitem{lykken-witten}
E.~Witten, \emph{Strong coupling expansion of Calabi-Yau
  compactification}, \npb{471}{1996}{135} [\hepth{9602070}];\\
J.D. Lykken, \emph{Weak scale superstrings}, \prd{54}{1996}{3693}
[\hepth{9603133}];\\
C.P.~Bachas, \emph{Desert in energy or transverse space?}, 
in \emph{Goeteborg 1998, novelties in string theory},
  pp.~287--293;
\emph{Unification with low string scale}, \jhep{11}{1998}{023}
[\hepph{9807415}].
%%CITATION = HEP-PH 9807415;%%.

\bibitem{polchinski}
J.~Polchinski, \emph{Dirichlet-branes and Ramond-Ramond charges},
\prl{75}{1995}{4724} [\hepth{9510017}].

\bibitem{AKT} 
See e.g.\ I.~Antoniadis, E.~Kiritsis and T.N. Tomaras, \emph{A D-brane
alternative to unification}, \plb{486}{2000}{186} [\hepph{0004214}];
\emph{D-brane standard model}, \forp{49}{2001}{573} [\hepth{0111269}].

\bibitem{AADD}
I.~Antoniadis, N.~Arkani-Hamed, S.~Dimopoulos and G.R. Dvali,
  \emph{New dimensions at a millimeter to a Fermi and superstrings at
  a TeV}, \plb{436}{1998}{257} [\hepph{9804398}];\\
I.~Antoniadis and C.~Bachas, \emph{Branes and the gauge hierarchy},
\plb{450}{1999}{83} [\hepth{9812093}].

\bibitem{ADD}
N.~Arkani-Hamed, S.~Dimopoulos and G.R. Dvali, \emph{The hierarchy
  problem and new dimensions at a millimeter}, \plb{429}{1998}{263}
  [\hepph{9803315}].

\bibitem{rs}
L. Randall and R. Sundrum, \emph{A large mass hierarchy from a small
  extra dimension}, \prl{83}{1999}{3370} [\hepph{9905221}];
\emph{An alternative to compactification}, \prl{83}{1999}{4690}
[\hepth{9906064}].

\bibitem{binetruy}
P.~Bin\'etruy, C.~Deffayet and D.~Langlois, \emph{Non-conventional
  cosmology from a brane-universe}, \npb{565}{2000}{269}
  [\hepth{9905012}];\\
P.~Bin\'etruy, C.~Deffayet, U.~Ellwanger and D.~Langlois, \emph{Brane
  cosmological evolution in a bulk with cosmological constant},
  \plb{477}{2000}{285} [\hepth{9910219}].

\bibitem{ida}
P.~Kraus, \emph{Dynamics of anti-de~Sitter domain walls},
\jhep{12}{1999}{011} [\hepth{9910149}];\\
D.~Ida, \emph{Brane-world cosmology}, \jhep{09}{2000}{014}
[\grqc{9912002}].

\bibitem{mirage}
A.~Kehagias and E.~Kiritsis, \emph{Mirage cosmology},
\jhep{11}{1999}{022} [\hepth{9910174}].

\bibitem{muko}
S.~Mukohyama, T.~Shiromizu and K.-i. Maeda, \emph{Global structure of
  exact cosmological solutions in the brane world},
  \prd{62}{2000}{024028} [\hepth{9912287}].

\bibitem{tym}
E.~Kiritsis, \emph{Supergravity, D-brane probes and thermal super
Yang-Mills: a comparison}, \jhep{10}{1999}{010} [\hepth{9906206}].

\bibitem{hall}
L.J. Hall and D.R. Smith, \emph{Cosmological constraints on theories
  with large extra dimensions}, \prd{60}{1999}{085008}
  [\hepph{9904267}];\\
S.~Hannestad, \emph{Strong constraint on large extra dimensions from
   cosmology}, \prd{64}{2001}{023515} [\hepph{0102290}].

\bibitem{hebecker}
C.~van de Bruck, M.~Dorca, C.J.~Martins and M.~Parry,
\emph{Cosmological consequences of the brane/bulk interaction},
\plb{495}{2000}{183} [\hepth{0009056}];\\
%%CITATION = HEP-TH 0009056;%%
U.~Ellwanger, \emph{Cosmological evolution in compactified
  Ho\v{r}ava-Witten theory induced by matter on the branes},
  \epjc{25}{2002}{157} [\hepth{0001126}];\\
A.~Hebecker and J.~March-Russell, \emph{Randall-sundrum ii cosmology,
AdS/CFT and the bulk black hole}, \npb{608}{2001}{375}
[\hepph{0103214}];\\
P.~Brax, C.~van de Bruck and A.C.~Davis,
\emph{Brane-world cosmology, bulk scalars and perturbations},
\jhep{10}{2001}{026} [\hepth{0108215}];
%%CITATION = HEP-TH 0108215;%%
D.~Langlois, L.~Sorbo and M.~Rodriguez-Martinez, \emph{Cosmology of a
  brane radiating gravitons into the extra dimension},
  \prl{89}{2002}{171301} [\hepth{0206146}].

\bibitem{ktt}
E.~Kiritsis, N.~Tetradis and T.N. Tomaras, \emph{Induced gravity on RS
  branes}, \jhep{03}{2002}{019} [\hepth{0202037}].

%\bibitem{steady}
%F. Hoyle and J.V. Narlikar, \emph{Time symmetric electrodynamics and
%the arrow of time in cosmology},\emph{Proc.\ Royal Soc.} {\bf A277}
%(1964) 1;
%\emph{Electrodynamics of direct interparticle action, I. The quantum
%mechanical response of the universe}, \ap{54}{1969}{207}.

%\bibitem{perl}
%{\sc Supernova Search Team} collaboration, A.G. Riess et al.,
%  \emph{Observational evidence from supernovae for an accelerating
%  universe and a cosmological constant}, \emph{Astron. J.} {\bf 116}
%  (1998) 1009 [\astroph{9805201}];\\
%{\sc Supernova Cosmology Project} collaboration, S.~Perlmutter et al.,
%  \emph{Measurements of omega and lambda from 42 high-redshift
%  supernovae}, \apj{517}{1999}{565} [\astroph{9812133}].

%\bibitem{kamke}
%E. Kamke, \emph{Differentialgleichungen L\"osungsmethoden und  
%L\"osungen}, Akademische Verlagsgesellschaft Geest und Portig K.-G., Leipzig 1961. 

\bibitem{tetra} 
P. Apostolopoulos and N. Tetradis, \emph{Brane Cosmology with Matter in the Bulk}, 
University of Athens preprint, 2004.

\end{thebibliography}
\end{document}